\documentstyle[12pt]{article}
\newlength{\mathspace}
\def\np#1{ Nucl. Phys. B#1}
\def\pr#1    { Phys. Rev. D#1 }
\def\pl#1{ Phys. Lett. B#1}

\def\ijmp#1  { Int. Jour. Mod. Phys. A#1 }
\def\mpl#1   { Mod. Phys. Lett. A#1 }
\def\begineq{\begin{equation}}
\def\endeq{\end{equation}}
\def\eqabegin{\begin{eqnarray}}
\def\eqaend{\end{eqnarray}}

\begin{document}
\baselineskip=0.7cm
\setlength{\mathspace}{2.5mm}



\begin{titlepage}

    \begin{normalsize}
     \begin{flushright}
  		 CTP-TAMU-29/98, SINP-TNP/98-17\\
                 hep-th/9807139\\
     \end{flushright}
    \end{normalsize}
    \begin{Large}
       \vspace{.4cm}
       \begin{center}
         {\bf 4-String Junction and Its Network}\\ 
       \end{center}
    \end{Large}

\begin{center}
           
             \vspace{.4cm}

            J. X. L{\sc u}$^1$\footnote[2]{ E-mail address: 
            jxlu@phys.physics.tamu.edu}  and   
            Shibaji R{\sc oy}$^2$\footnote{E-mail address: 
            roy@tnp.saha.ernet.in} 

            \vspace{2mm}

            $^1${\it Center for Theoretical Physics, Physics Department}\\
            {\it Texas A \& M University, College Station, TX 77843, USA}\\

            \vspace{2mm}

        $^2${\it Saha Institute of Nuclear Physics, 1/AF 
Bidhannagar, Calcutta 700 064, India}\\

      \vspace{.6cm}

    \begin{large} ABSTRACT \end{large}
        \par
\end{center}
We study a BPS configuration in which four strings (of different type) meet at 
a point in $N = 2, D = 8$ supergravity, i.e., the low energy effective theory
of $T^2$-compactified type II string theory. We demonstrate that the charge 
conservation of the four strings implies the vanishing of the net force
(due to the tensions of various strings) at the 
junction and vice versa, using the tension formula for
$SL(3, Z)$ strings obtained recently by the present authors. We then show that 
a general 4-string junction preserves 1/8 of the spacetime supersymmetries. 
Using 4-string junctions as building blocks, we construct a string 
network which also preserves 1/8 of the spacetime supersymmetries.
\end{titlepage}
\vfil\eject

Type IIB string theory admits a stable BPS configuration in which three strings
of different type meet at a point, i.e., a 3-string junction [1,2,3,4]. The 
existence of such a 3-string junction is due to the D-string picture [5] 
and the 
special properties of electrodynamics in two dimensions as well as the 
strong-weak U-duality $SL(2,Z)$ symmetry. D-string picture allows a F-string, 
i.e., (1,0)-string, 
ending on a D-string, i.e., (0,1)-string. The consistent coupling at the 
endpoint guarantees the
charge conservation [6]. The charge at the endpoint of the F-string appears as 
a point source to the U(1) field on the D-string worldsheet. The Gauss law in 
one spatial dimension implies that the field strength $F_{01}$ can be chosen as
the constant charge flux on one side of the point charge along the D-string 
and zero on the other side [4]. This can also be understood from the picture 
that the other end of the F-string may be thought to end on the D-string at 
infinity, implying that the charge flux goes along the D-string 
only from the positive charge 
endpoint to the negative charge endpoint of the F-string at infinity.  
Due to this, one side of the original D-string can no longer be a D-string 
but a (1,1)-string (or ($-1$,$-1$)-string depending on
the orientation) and the charge at the endpoint of F-string causes one half of 
the D-string to bend rigidly rather than to be spike-like as in the case for 
D p-brane for $p > 2$ [7], therefore giving rise to a stable 3-string junction. 
The U-duality symmetry $SL(2, Z)$ guarantees the existence of a general 
$(p,q)$-string with coprime integers $p$ and $q$. Otherwise, the 3-string 
junction would not be possible. As an interesting application the 3-string
junction has been used in [8] to construct the 1/4 BPS states in $D = 4$,
$N = 4$ supersymmetric Yang-Mills theory.  Also, the dynamics of 3-string
junction has been studied in [9,10] and the M-theoretic origin of such
junctions has been pointed out in [1,11,12].

The $T^2$-compactified Type II string theory admits a general stable BPS
$(q_1, q_2, q_3)$-string configuration with any two of 
the three integers $q_1,
q_2$ and $q_3$ being coprime [13]. All possible $(q_1, q_2, q_3)$ strings fill 
in the various multiplets of the strong-weak symmetry $SL(3,Z)$ of the complete 
U-duality symmetry $SL(3,Z) \times SL(2,Z)$ in this
theory (the strings are inert under the T-dual symmetry $SL(2,Z)$) 
and each of the strings breaks half of the spacetime supersymmetries. 
The three
integral charges $q_i$ are associated with the corresponding 
three 2-form potentials $B^{(i)}_{\mu\nu}$ ($i = 1, 2, 3$) which form  a 
triplet of the $SL(3,Z)$ in this theory. We therefore expect that we have three 
kinds of strings. In particular, the $(1,0,0)$-string is the $T^2$-compactified 
Type IIB F-string and the $(0,1,0)$-string is the 
$T^2$-compactified D-string while the $(0, 0, 1)$-string is the 
wrapped D 3-brane. If we draw the analogy between what we have here 
and what has been discussed in the previous paragraph for the 3-string 
junction, one natural conclusion is that we should have a 4-string junction
in $D = 8$. The only thing which we have not explained well is how the $(1,0,0)$
and $(0,1,0)$ strings couple consistently with the $(0,0,1)$-string so that
one can have a 4-string junction. This can be explained if we trace 
the origin of these
strings back to $D = 10$. In other words, in order to have a stable BPS 
4-string junction in $D = 8$, we should have a F-string and a D-string ending 
consistently on a D 3-brane in $D = 10$. This possibility certainly exists if 
one examines the effective D 3-brane worldvolume action [14,15] in which 
both the NSNS and RR 2-form potentials couple consistently with 
 the U(1) gauge field on the worldvolume. Thanks to the
recent work [16] where such a stable BPS configuration has been 
discovered explicitly
and is shown to preserve 1/4 of the D 3-brane worldvolume supersymmetries. In 
general, a toroidal compactification on either supergravity theory or super
p-brane action preserves the original supersymmetries. Loosely speaking,
the 4-string junction can be viewed as the $T^2$-dimensionally reduced  
$D = 10$ BPS configuration of
a F-string and a D-string ending on a D 3-brane as we will discuss in the 
following.
  This further convinces us that there should exist a 4-string junction in 
$D = 8$ which preserves 1/8 of the spacetime supersymmetries. 

Let us explain further the 4-string junction from $D = 10$ picture. 
We start with  a F-string and a D-string ending on a D 3-brane.
To be specific, let us assume
that  the D 3-brane be along $x^3, x^8$ and $x^9$ directions, the F-string be 
along the $x^1$-direction with its endpoint on the D 3-brane at 
$x^3 = 0, x^8 = a, x^9 = 0$, and the D-string along the 
$x^2$-direction with its endpoint on the D 3-brane at 
$x^3 = 0, x^8 = 0, x^9 = b$ where $a$ and $b$ are two constants. If
such a configuration is a stable BPS one, both the F-string and D-string are 
spike-like [16,7]. The charge at the F-string endpoint and the
charge at the D-string endpoint appear to be electric and magnetic, 
respectively, to the U(1) field on the D 3-brane worldvolume. 

Now let us compactify the
coordinates $x^8$ and $x^9$ on a torus and insist that the 3-brane wrap on 
the torus. If we shrink the spacetime torus to zero,  the 3-brane then becomes 
the (0,0,1)-string along $x^3$ in $D = 8$. At the same time, the F-string and
D-string become (1,0,0)-string and (0,1,0)-string, respectively, in $D = 8$.
Their endpoints must meet at $x^3 = 0$ ending on the (0,0,1)-string.
The $D = 10$ type IIB NSNS and RR 2-form potentials are reduced to the 
$D = 8$ 2-form potentials $B^{(1)}_2$ and $B^{(2)}_2$, respectively. The U(1)
gauge field on the D 3-brane worldvolume is reduced to the U(1) gauge field 
on the (0,0,1)-string worldsheet with the only field strength component 
$F_{03}$.

 On reducing the D 3-brane worldvolume action to give the action
describing the (0,0,1)-string in $D = 8$, we know that $F_{03}$ couples to
both $B^{(1)}_2$ and $B^{(2)}_2$. We also know that $B^{(1)}_2$ and 
$B^{(2)}_2$
determine the charges for (1,0,0)-string and (0,1,0)-string, respectively, 
in $D = 8$. So in $D = 8$, both the (1,0,0)-string and (0,1,0)-string 
endpoint charges provide point sources to the U(1) field strength $F_{03}$
on the (0,0,1)-string worldsheet (the two charges are now located at the 
same point $x^3 = 0$)\footnote{In two dimensions, we do not have a 
distinction between electric and magnetic charges. But in the present 
case, we can still make a distinction between the two charges since their 
dependences on the string coupling are different when they act as  sources to
 $F_{03}$.}. 

With a similar discussion as we did above for the 3-string 
junction, we must conclude that the (0,0,1)-string cannot remain as it is when
a (1,0,0)-string and a (0,1,0)-string end on the (0,0,1)-string at $x^3 =0$. 
The two sides of the (0,0,1)-string around $x^3 = 0$ must bend rigidly according
to the following two cases: 1) The $x^3 > 0$ (or $x^3 < 0$) side becomes 
(1,1,1)-string (or ($-1$,$-1$,$-1$)-string depending on the oritentation) 
while the other side remains as (0,0,1)-string if the two other 
ends of the (1,0,0) and
(0,1,0) strings meet with the (0,0,1)-string at infinity on $x^3 > 0$ (or 
$x^3 < 0$) side. 2) The $x^3 > 0$ (or $x^3 < 0$) side becomes (1,0,1)-string and
the other side becomes (0,1,1)-string if the other end of the (1,0,0)-string
meets with the (0,0,1)-string at infinity on $x^3 > 0$ (or $x^3 < 0$) side while
that of the (0,1,0)-string meets with the (0,0,1)-string at infinity 
on $x^3 < 0$ (or $x^3 > 0$) side. In each case, we have a 4-string junction.

The above discussion clearly indicates that there should exist a stable BPS 
4-string junction in $D = 8$. However, at this point, the only thing that we 
are certain about is that the existence of 
various couplings indeed allows 4 strings
in $D = 8$ to meet at one point. In other words, the charge conservation is
not violated when 4 strings meet. Whether there actually exists a stable BPS
4-string junction requires us to show that the net-force due to string tensions
at the junction should vanish and the junction should preserve some unbroken
supersymmetries. We will show that the answers for both of these are positive.

If the four strings are of type $(q_1^{(i)}, q_2^{(i)}, q_3^{(i)})$ 
$(1 \leq i \leq 4)$,
then the charge conservation [13,6,15,17] implies 
\begineq
\sum_{i = 1}^4 q_1^{(i)} = \sum_{i = 1}^4 q_2^{(i)} = \sum_{i = 1}^4 q_3^{(i)}
= 0.
\endeq
We recall from [13] that the tension for a $(q_1, q_2, q_3)$-string in string 
metric is
\begineq
T_{(q_1,q_2,q_3)} = \sqrt{e^{-\phi_0 + {\sqrt 3} \varphi_0} q_3^2 + 
e^{- 2\phi_0} (q_2 - \chi_{10} q_3)^2 + (q_1 - \chi_{30} q_2 - \chi_{20} q_3)^2
}, 
\endeq
where $\phi_0, \varphi_0, \chi_{10}, \chi_{20}$ and $\chi_{30}$ are the 
asymptotic values of the scalars parametrizing the coset $SL(3,R)/SO(3)$ in
the $D = 8, N = 2$ supergravity. In particular, $\phi$ and $\chi_3$ are the 
$D = 10$ type IIB dilaton and RR scalar, respectively. $\varphi$ is another 
dilatonic scalar while $\chi_1$ and $\chi_2$ are axions, all of which are due to
the $T^2$-compactification of type IIB supergravity. 
The above tension can be viewed as the magnitude of the following tension vector
\begineq
\vec{T}_{(q_1, q_2, q_3)} = (q_1 - \chi_{30} q_2 - \chi_{20} q_3)\, \hat{e}_1 
                            + e^{- \phi_0} (q_2 - \chi_{10} q_3)\, \hat{e}_2
                    + e^{-\phi_0/2 + {\sqrt 3} \varphi_0 /2} q_3\, \hat{e}_3,
\endeq
where $\hat{e}_i$ ($i = 1, 2, 3$) are the unit vectors along the 
$x^i$-directions, respectively. This corresponds to orienting the 
$(q_1, q_2, q_3)$-string along the direction of the tension vector.
 We use two simple examples to justify
the above tension vector. First,  the (1,0,0), (0,1,0) and (0,0,1) strings 
obtained from a F-string and a D-string ending on a D 3-brane by shrinking 
the spacetime torus to zero-size are perpendicular to each other (the axions are
all set to zero). The above tension vector formula indeed shows this feature. 
Secondly, as discussed in [13], our tension formula reduces to the one for the 
type IIB $(q_1, q_2)$-string when $\varphi, \chi_1, \chi_2$ and $q_3$ are
all set to zero. In this case, our above tension vector goes over to the one 
given by Sen [18] in discussing the 3-string junction network. 

  With the above tension vector, we have the zero net force automatically at
the 4-string junction
\begineq
\sum_{i = 1}^4 \vec{T}_{(q_1^{(i)}, q_2^{(i)}, q_3^{(i)})} = 0,
\endeq
provided the charge conservation (1) is satisfied. The converse is also true, 
i.e.,  the zero net force at the 4-string junction requires the charge 
conservation of eq.(1).  This beautiful property further convinces us that
the tension vector (3) is indeed correct.

We shall now show that a 4-string junction preserves 1/8 of the spacetime
supersymmetries as expected. Before we do so, let us seek first the 1/2-unbroken
supersymmetry condition for a stable BPS (1,0,0)-string configuration of 
the $D = 8, N = 2$ supergravity. This SUSY condition can be obtained from that 
of the F-string or (1,0)-string configuration in $D = 10$ type IIB 
supergravity. Let $\epsilon_L$ and $\epsilon_R$ be the two real 
Majorana-Weyl supersymmetry transformation parameters of type IIB string 
theory, associated with the left and the right moving sector of the worldsheet
of the F-string. Assuming the F-string to be along $x^1$, we have the SUSY 
condition for the (1,0)-string configuration as\footnote{Here and in the 
subsequent discussion, spinors in the corresponding SUSY condition all take 
their asymptotic values.}
\begineq
\epsilon_L + i \epsilon_R = \Gamma^2 \Gamma^3 \cdots \Gamma^9 (\epsilon_L - i
\epsilon_R),
\endeq
where $\Gamma^M$ ($M = 0, 1, \cdots 9$) are the ten dimensional gamma matrices.
The convenient representation for these matrices is the Majorana one in which
$\Gamma^0$ is real and antisymmetric while the rest are real and symmetric.
A concrete construction for the $\Gamma^M$ is given in the appendix. In this 
representation, the above condition is equivalent to 
\begineq
\epsilon_L = \Gamma^2 \Gamma^3 \cdots \Gamma^9 \epsilon_L,\qquad
\epsilon_R = - \Gamma^2 \Gamma^3 \cdots \Gamma^9  \epsilon_R.
\endeq
If we are able to express the $\epsilon_L$ and $\epsilon_R$ in terms of the
two pseudo Majorana supersymmetry transformation parameters $\epsilon^{(1)}$
and $\epsilon^{(2)}$ in $D = 8, N =2$ supergravity and express $\Gamma^M$
in terms of $D = 8$ gamma matrices $\gamma^\mu$, then the above SUSY conditon 
is reduced to the one for the $(1,0,0)$-string configuration (this string is
also along $x^1$). 
This has been done in detail in the appendix and the required 
SUSY condition is
\begineq
\left(\begin{array}{c}
       \epsilon^{(1)}\\
       \epsilon^{(2)}\end{array}\right) = \gamma^0 \gamma^1 \otimes \sigma_1
  \left(\begin{array}{c}
       \epsilon^{(1)}\\
       \epsilon^{(2)}\end{array}\right),
\endeq
where $\sigma_1$ is the $2\times 2$ Pauli matrix.
  
As is well known in supergravity theories, the spinors are singlet for the 
non-compact Cremmer-Julia symmetry $G$ but form certain representation of the
local hidden symmetry $H$ which is isomorphic to the maximal compact subgroup
of $G$ linearly realized on the physical states. In generating a general 
U-duality p-brane solution from a single charge solution as we did in [13],
all we need to care about is the bosonic fields since the fermionic fields
are always set to zero. Therefore, we do not need to use the 
``vielbein"-formalism
for the scalar fields. In other words, we need to employ only the non-compact 
global Cremmer-Julia symmetry $G$ to find the general U-duality p-brane 
solution. By this, we concluded in [13] that the general U-duality p-brane 
solution should preserve the same number of unbroken supersymmetries as that
for the original single charge solution. Here the story is different. We no 
longer consider an isolated p-brane solution but a few strings meet at a point.
Since the strings in a 4-string junction share the same $SL(3,R)/SO(3)$ moduli
but different charges, they are physically inequivalent. Each of these strings
can be obtained through a different global $SL(3,R)$ transformations.
If we choose to work
in the ``vielbein"-formalism and in a unitary gauge, then an induced 
SU(2) ($\simeq SO(3)$) transformation will arise whenever a $SL(3,R)$ 
transformation 
 is performed [19,20]. This induced SU(2) transformation will act on the
pseudo Majorana spinors $\epsilon^{(1)}$ and $\epsilon^{(2)}$ which form a 
doublet of SU(2). So the SUSY condition for each of the 4 strings in 
a 4-string junction will be differnt. We will just make use of such 
differences to determine if any supersymmetry is preserved for
a 4-string junction. 

Since $\left(\begin{array}{c} \epsilon^{(1)}\\  \epsilon^{(2)}\end{array}
\right)$ is a doublet of SU(2), the SUSY condition (7) for the (1,0,0)-string 
configuration must be a special form of a general SU(2) covariant expression of 
a general $(q_1, q_2, q_3)$-string since they both preserve 1/2 of the spacetime
supersymmetries and are related to each other by an induced  SU(2) 
transformation. Examing the SUSY condition (7) for the (1,0,0)-string, 
we note that we already have
an appearance of the SU(2) generator $\sigma_1$. Our hunch for the 
covariant SUSY condition for a general $(q_1, q_2, q_3)$-string along the
$x^1$ axis is\footnote{
Similar SUSY condition was also given in a recent paper by Bhattacharya et al 
[21] in discussing the non-planar network of 3-string junctions. We, however, 
differ from theirs in the following ways. They used the product of gamma 
matrices in the transverse directions of the string in their SUSY condition 
while we have here the product of gamma matrices along the longitudinal 
directions of the string. The two are different since in D = 8, the pseudo 
Majorana spinors cannot be Weyl at the same time. The subsequent discussions are
also differnt. They discussed the nonplanar network of 3-string junctions while
we are here discussing a 4-string junction and its network.}
\begineq
\left(\begin{array}{c} 
\epsilon^{(1)}\\  
\epsilon^{(2)}
\end{array}\right) = \gamma^0 \gamma^1 \otimes \frac{V}{|V|} 
\left(\begin{array}{c} 
\epsilon^{(1)}\\  
\epsilon^{(2)}\end{array} \right),
\endeq
where $V = V_a \sigma_a$ with $V_a$ a SO(3) vector and 
$\sigma_a$ ($a = 1, 2, 3$) the three Pauli matrices. $|V|$ denotes the 
magnitude of $V$. Note that $V_a$ 
 depend only on the asymptotic values of $SL(3,R)/SO(3)$ moduli and 
the three charges $q_1, q_2, q_3$. 
Whether we succeed in reaching the above condition depends on if we can find
the vector $V_a$. 

We recall from [13] that the scalar matrix ${\cal M}_3$ 
parametrizing the coset $SL(3,R)/SO(3)$ is
\begineq
{\cal M}_3 = e^{ \frac{\varphi}{\sqrt 3}}\left(\begin{array}{ccc}
e^{-\phi} + \chi_3^2 e^{\phi}  &  \chi_3 e^{\phi}  
&(\chi_2 + \chi_1 \chi_3)e^{-{\sqrt 3} \varphi}\\
+ (\chi_2 + \chi_1 \chi_3)^2 e^{-{\sqrt 3} \varphi}&+\chi_1 (\chi_2 + 
\chi_1 \chi_3) e^{-{\sqrt 3} \varphi}&\\
&&\\
\chi_3 e^{\phi} & e^{\phi} + \chi_1^2 e^{-{\sqrt 3} \varphi}&
\chi_1 e^{-{\sqrt 3} \varphi}\\
+ \chi_1 (\chi_2 + \chi_1 \chi_3)e^{-{\sqrt 3} \varphi}&&\\
&&\\
(\chi_2 + \chi_1 \chi_3)e^{-{\sqrt 3} \varphi}&\chi_1 e^{-{\sqrt 3} \varphi}
&e^{-{\sqrt 3} \varphi}\end{array}\right).
\endeq

The matrix ${\cal M}_3$ can be re-expressed in a 3-bein form as
${\cal M}_3 = \nu \nu^T$ with the 3-bein $\nu_{ia}$. Here $i = 1, 2, 3$ are
 the $SL(3,R)$ indices while $a = 1, 2, 3$ are the $SO(3)$ indices. 
The 3-bein $\nu$ transforms under
the global $SL(3,R)$  and local $SO(3)$ as $\nu \rightarrow \Lambda \nu R$, 
with
$\Lambda$ a global $SL(3,R)$ matrix and $R$ a $SO(3)$ rotation matrix. 
We also know that the charge triplet 
\begineq
q = \left(\begin{array}{c}
q_1\\
q_2\\
q_3 \end{array}\right)
\endeq
transforms only under $SL(3,R)$ as $q \rightarrow \Lambda q$. So the
$SO(3)$ vector $V$ can be constructed as $V_a = (\nu^{-1}_0)_{ai} q_i$ with
\begineq
\nu^{-1} = e^{- \varphi / 2{\sqrt 3}}\left(\begin{array}{ccc}
e^{\phi/2} & - \chi_3 e^{\phi/2} & - \chi_2 e^{\phi/2}\\
0 & e^{- \phi/2} & - \chi_1 e^{- \phi/2}\\
0 & 0 & e^{{\sqrt 3} \varphi /2} \end{array} \right).
\endeq
The subscirpt `0' always means that the scalars take their asymptotic values.
The vector $V_a$ transforms only under SO(3) as $V_a \rightarrow R_{ba} V_b$.
The $SO(3)$ action on the vector $V_a$ can also be realized on a $SU(2)$ basis
through $V = V_a \sigma_a$ as $V \rightarrow U^+ V U$, with
\begineq
U^+ \sigma_a U = R_{ab} \sigma_b,
\endeq
where $U^+ = U^{-1}$ with $U$ a SU(2) matrix.

Therefore, it can be shown that eq. (8) is indeed covariant under a 
SU(2) transformation U
if the spinor doublet transforms as
\begineq
\left(\begin{array}{c} 
\epsilon^{(1)}\\  
\epsilon^{(2)}
\end{array}\right) \rightarrow U^{-1} \left(\begin{array}{c} 
\epsilon^{(1)}\\  
\epsilon^{(2)}
\end{array}\right).
\endeq

 So far in discussing the SUSY condition (8), we always assume that the 
($q_1, q_2, q_3$)-string be along the direction of $x^1$.
 We could not have found anything wrong with this assumption if we had 
never discussed various string junctions. We recall that the zero net force at
a 4-string junction requires the orientation of the ($q_1, q_2, q_3$)-string
 be given by the tension vector (3). This implies that except for a 
($q_1, 0, 0$)-string, a general ($q_1, q_2, q_3$)-string cannot be kept along
the $x^1$-direction. Therefore, the superscript 1 in the $\gamma^1$-matrix 
in the SUSY condition (8) should be replaced by a label $1'$ characterizing 
the direction of the tension vector (3). We have
\begineq
\gamma^{1'} = \sin \alpha  \cos \beta \,\gamma^1 + \sin \alpha \sin \beta \,
\gamma^2 + \cos \alpha\, \gamma^3,
\endeq
where $\sin \alpha, \cos \alpha, \sin \beta, \cos \beta$ can be 
determined by the tension vector ${\vec T}_{(q_1, q_2, q_3)}$ as
\eqabegin
\sin \alpha \cos \beta &=& \frac{q_1 - \chi_{30} q_2 - \chi_{20} q_3}{T_{(q_1,
q_2, q_3)}},\nonumber\\
\sin \alpha \sin \beta &=& \frac{e^{-\phi_0}\, (q_2 - \chi_{10}q_3)} 
{T_{(q_1,
q_2, q_3)}},\nonumber\\
\cos \alpha &=& \frac{e^{- \phi_0/2 + {\sqrt 3} \varphi_0 /2}\, q_3}{T_{(q_1,
q_2, q_3)}},
\eqaend
where the tension $T_{(q_1, q_2, q_3)}$ is given by eq. (2).
The tension vector can now be re-expressed as
\begineq
{\vec T}_{(q_1, q_2, q_3)} = T_{(q_1, q_2, q_3)} (\sin \alpha  \cos \beta 
\,{\hat e}_1 + \sin \alpha \sin \beta\, {\hat e}_2
+ \cos \alpha \,{\hat e}_3).
\endeq

If we introduce three Cartesian unit vectors ${\hat u}_a$ for the internal 
three dimensional tangent space of the coset $SL(3,R)/SO(3)$ space, 
then we can express $V_a$ in a vector form
as ${\vec V} = V_a {\hat u}_a$. Now if we employ eq. (11) to express 
the components 
$V_a = (\nu_0^{-1})_{ai} q_i$ explicitly, we have, using eq. (15), 
\begineq
{\vec V} = |V| (\sin \alpha  \cos \beta 
\,{\hat u}_1 + \sin \alpha \sin \beta\, {\hat u}_2
+ \cos \alpha \,{\hat u}_3),
\endeq
where $|V|$, the magnitude of $\vec V$ or $V = V_a \sigma_a$, is actually the 
($q_1, q_2, q_3$)-string tension written in Einstein metric [13]. 
Comparing eq. (16)
with the above equation, we can easily see that  
the tension vector re-expressed in 
Einstein metric and the vector ${\vec V}$ behave in exactly the same way. 
The only difference is that they are expressed in two different spaces,
namely, the tension vector is in three-dimensional space-time coordinate
system whereas, the vector ${\vec V}$ is in the three-dimensional tangent
space of the coset space $SL(3,R)/SO(3)$. 
Since both 
${\vec T}_{(q_1,q_2,q_3)}$ and ${\vec V}$ are defined for arbitrary asymptotic 
values of the moduli and the three charges $q_1, q_2, q_3$, it 
would be strange if the two three-dimensional spaces are not identified given 
the above behavior even though we do not have to do so for the present purpose.
The identification of these two spaces was also made in [21].
The similar behavior also occurs for a general type IIB $(p,q)$-string 
and the
corresponding two 2-dimensional spaces were identified in the discussion of 
the planar string network in [18].  There must exist a deep reason for 
such an identification of two completely different spaces beyond what
we have explained here. This may be connected to the well-known fact that 
there is a relationship between the unbroken SUSY condition and the so-called 
no-force condition.

Now we finally come to show that a general 4-string junction 
preserves 1/8 of the 
spacetime supersymmetries. Let us show this first for the simplest 4-string
junction which consists of $(1, 0, 0), (0, 1, 0), (0, 0, 1)$ and $(-1, -1, -1)$
strings with $\chi_{10} = \chi_{20} = \chi_{30} = 0$. The SUSY condition for 
the $(1,0,0)$-string is, with $\gamma^{1'} = \gamma^1$ and $V/|V| = \sigma_1$,
\begineq
\left(\begin{array}{c} 
\epsilon^{(1)}\\  
\epsilon^{(2)}
\end{array}\right) = \gamma^0 \gamma^1 \otimes \sigma_1 \left(\begin{array}{c} 
\epsilon^{(1)}\\  
\epsilon^{(2)}
\end{array}\right),
\endeq
which is equivalent to $\epsilon^{(1)} = \gamma^0 \gamma^1 \epsilon^{(2)}$. 
Since $(\gamma^0 \gamma^1)^2 = 1$, so one half of the spacetime supersymmeties
is preseved by the $(1,0,0)$-string configuration. For the $(0,1,0)$-string, 
we have, $\gamma^{1'} = \gamma^2$ and $V/|V| = \sigma_2$. So the SUSY condition
becomes,
\begineq
\left(\begin{array}{c} 
\epsilon^{(1)}\\  
\epsilon^{(2)}
\end{array}\right) = \gamma^0 \gamma^2 \otimes \sigma_2 \left(\begin{array}{c} 
\epsilon^{(1)}\\  
\epsilon^{(2)}
\end{array}\right),
\endeq
which is independent of the SUSY condition for the $(1, 0,0)$-string. The
above is equivalent to $\epsilon^{(1)} = - i \gamma^0 \gamma^2 \epsilon^{(2)}$.
The above two independent SUSY conditions preserve 1/4 of the spacetime 
supersymmetries. For the $(0,0,1)$-string, we have,  
$\gamma^{1'} =\gamma^3$ and $V/|V| = \sigma_3$. The SUSY condition in this case
has the form,
\begineq
 \left(\begin{array}{c} 
\epsilon^{(1)}\\  
\epsilon^{(2)}
\end{array}\right) = \gamma^0 \gamma^3 \otimes \sigma_3 \left(\begin{array}{c} 
\epsilon^{(1)}\\  
\epsilon^{(2)}
\end{array}\right),
\endeq
which is equivalent to $\epsilon^{(1)} = \gamma^0 \gamma^3 \epsilon^{(1)}$
and $\epsilon^{(2)} = - \gamma^0 \gamma^3 \epsilon^{(2)}$. This breaks further
1/2 of what is left, i.e., preserving 1/8 of the spacetime supersymmetries.
Does the $(-1, -1, -1)$-string break more supersymmetries? The SUSY condition 
for this string is automatically satisfied once we have all the above three 
conditions. In showing this, we need also the following deduced relations
\eqabegin
\gamma^1 \otimes \sigma_2 \left(\begin{array}{c} 
\epsilon^{(1)}\\  
\epsilon^{(2)}
\end{array}\right) &=&- \gamma^2 \otimes \sigma_1  \left(\begin{array}{c} 
\epsilon^{(1)}\\  
\epsilon^{(2)}
\end{array}\right),  \nonumber\\
\gamma^1 \otimes \sigma_3 \left(\begin{array}{c} 
\epsilon^{(1)}\\  
\epsilon^{(2)}
\end{array}\right) & =& - \gamma^3 \otimes \sigma_1 \left(\begin{array}{c} 
\epsilon^{(1)}\\  
\epsilon^{(2)}
\end{array}\right),\nonumber\\
\gamma^2 \otimes \sigma_3 \left(\begin{array}{c} 
\epsilon^{(1)}\\  
\epsilon^{(2)}
\end{array}\right)&=& - \gamma^3\otimes\sigma_2 \left(\begin{array}{c} 
\epsilon^{(1)}\\  
\epsilon^{(2)}
\end{array}\right).
\eqaend

In summary, the above 4-string junction indeed preserves 1/8 of the
spacetime supersymmetries. The needed SUSY conditions are summarized here as
\begineq
\left(\begin{array}{c} 
\epsilon^{(1)}\\  
\epsilon^{(2)}
\end{array}\right) = \gamma^0 \gamma^1\otimes \sigma_1 \left(\begin{array}{c} 
\epsilon^{(1)}\\  
\epsilon^{(2)}
\end{array}\right) = \gamma^0 \gamma^2 \otimes \sigma_2 \left(\begin{array}{c} 
\epsilon^{(1)}\\  
\epsilon^{(2)}
\end{array}\right) = \gamma^0 \gamma^3 \otimes \sigma_3 \left(\begin{array}{c} 
\epsilon^{(1)}\\  
\epsilon^{(2)}
\end{array}\right).
\endeq

One can show that the SUSY conditions for a general 4-string junction  are 
exactly the ones given in eq. (22). One way to check this is 
to note that the
three equations in (22) guarantee the SUSY condition to be satisfed for a 
general 
$(q_1, q_2, q_3)$-string with arbitrary asymptotic values of the moduli.
Therefore, a general 4-string junction satisfying charge conservation (1) and 
zero net-force condition (4) always preserves 1/8 of the spacetime 
supersymmetries. So a 4-string junction can be a stable BPS configuration.

Our discussion for 4-string junctions also provides basis for the corresponding
network in the same spirit of the planar string network discussed by Sen [18]. 
In the 4-string junction network, each 4-string junction satisfies the 
charge conservation (1) and 
zero net force condition (4), and preserves 1/8 
of the spacetime supersymmetries.
Since the SUSY conditions (22) are 
 independent of the composition of the network,
the 4-string junction network preserves also 1/8 of the spacetime 
supersymmetries. The orientations of links in the network are completely 
determined by the corresponding charge triplets associated with these links. 
Other topics such as string lattice and the compactification 
related to the string
network can also be discussed in the similar spirit as in [18].

With this work, we may expect that there exist other kinds of stable 
BPS string junctions 
(with maximum number of strings allowable by the corresponding U-duality 
symmetry and charge conservation in each junction) 
and the corresponding network in $D \le 7$. However, it appears that 
such a string junction and the corresponding network will exist
only in $D = 7$ i.e. 
$T^3$-compactified type II string theory and not in $D < 7$. 
In the case of $D = 7$ we expect that there exists
a 6-string junction. Both the 6-string junctions and the 
corresponding network preserves 1/32 of the spacetime supersymmetries. 
However, for $D < 7$, since the U-duality group becomes larger,
the corresponding vector representation becomes bigger than the 
spatial dimensions of the theory. It also appears that no supersymmetries can 
be preseved for those junctions. Thus in those cases, it is not
possible to form stable string junctions in a consistent fashion.
 
In conclusion, we have studied in this paper the BPS property of 
4-string junction in $N = 2$, $D = 8$ supergravity which is the
low energy effective theory of type II string theory compactified
on $T^2$. The existence of 4-string junction in $D = 8$ has been
argued to follow from the consistent coupling of F- and D-string
to the D 3-brane of type IIB theory in $D = 10$. We have shown 
that the junction  is stable since the net force due to the tensions
of various strings at the junction vanishes as a consequence of
charge conservation and vice-versa and the configuration preserves
1/8 of the space-time supersymmetries. We have also indicated how using
the 4-string junction as the building block one can construct a string
network and lattices. String junctions and the
corresponding networks are stable but curious states that exist in
string theory in various dimensions; however, their true utility either
in the formulation of non-perturbative string field theory [3]
or in the study of black holes [1,8] remains to be seen.

\vfil\eject
\begin{Large}
\begin{center}
{\bf Appendix}
\end{center}
\end{Large}
In this appendix, we will demonstrate how to express the real Majorana-Weyl 
spinors $\epsilon_L$ and $\epsilon_R$ in $D = 10$ in terms of two pseudo 
Majorana spinors $\epsilon^{(1)}$ and $\epsilon^{(2)}$ in $D = 8$ and how to 
obtain eq. (7) from eq. (5). The spacetime
signature is always chosen as $(-, + + \cdots +)$.

The eight $16 \times 16$ Dirac matrices $\Sigma^i$ $(i = 1, 2, \cdots, 8$) 
associated with $SO(8)$ can 
all be chosen as real and symmetric. An explicit representation for these 
$\Sigma^i$ can be found, for example, on page 288 in [22]. These matrices 
satisfy
the following Clifford algebra
\begineq
\{\Sigma^i, \Sigma^j\} = 2 \delta^{ij}.
\endeq
The corresponding $\Sigma^9$ can be defined as
\begineq
\Sigma^9 = \Sigma^1 \Sigma^2 \cdots \Sigma^8,
\endeq
which is real and symmetric and whose square is a unit matrix, i.e., 
$(\Sigma^9)^2 = 1$.
 
The SO(1,7) gamma matrices can therefore be defined as
\eqabegin
\gamma^0 &=& i \,\Sigma^1,\nonumber\\
\gamma^{j - 1} &=& \Sigma^j, \qquad (j = 2, 3, \cdots, 8).
\eqaend

In $D = 8$, one can show that there exist only either Weyl or pseudo Majorana
spinors. If we choose to work in the above representation 
for the gamma matrices,
a pseudo Majorana spinor $\eta$ is defined as
\begineq
\gamma^0 \eta^* = \eta,
\endeq
where $*$ denotes the complex conjugate. If we write 
$\eta =\lambda + i \zeta$ with $\lambda$ 
and $\zeta$ as two real spinors, 
then the above equation says $\zeta = - i \gamma^0 \lambda$,
so a general pseudo Majorana spinor in $D = 8$ can be written as
\begineq
\eta = (1 + \gamma^0) \lambda,
\endeq
with $\lambda$ an arbitrary real spinor in $D = 8$. Or we can express 
\begineq
\lambda = \frac{1}{2} (1 - \gamma^0) \eta.
\endeq

As is well-known, a Majorana-Weyl spinor $\Psi$ in $D = 10$ is real in a 
Majorana representation in which the gamma matrix $\Gamma^0$ is real and 
antisymmetric while the rest gamma matrices are real and symmetric. We can 
construct such a representation for $\Gamma^M$ ($M = 0, 1, 2, \cdots, 9$), 
which is useful for the purpose of this paper, as
\eqabegin
\Gamma^0 & = & - i\, \Sigma^1 \otimes \sigma_2,\nonumber\\
\Gamma^{j - 1} &=&  \Sigma^j \otimes I_2, \qquad (j = 2, \cdots, 9),\nonumber\\
\Gamma^9 &=& \Sigma^1 \otimes \sigma_3,
\eqaend
where $\Sigma^9$ is defined in eq. (24), $I_2$ is the $2\times 2$ unit matrix and
$\sigma_2$ and $\sigma_3$ are the usual Pauli matrices whose explicit forms are
\begineq
\sigma_1 = \left(\begin{array}{cc}
                 0&1\\
                 1&0\end{array}\right),\qquad \sigma_2 = \left(\begin{array}{cc}
                 0& - i\\
                 i&0\end{array}\right), \qquad 
\sigma_3 = \left(\begin{array}{cc}
                 1&0\\
                 0& -1 \end{array}\right).
\endeq
The corresponding $\Gamma_{11}$ in this representation is
\eqabegin
\Gamma_{11} &=& \Gamma^0 \Gamma^1 \cdots \Gamma^9,\nonumber\\
&=& \Sigma^1 \otimes \sigma_1,\nonumber\\
&=& \left(\begin{array}{cc}
                 0& \Sigma^1\\
                 \Sigma^1 &0\end{array}\right),
\eqaend
which is real and symmetric and $\Gamma_{11}^2 = 1$. A Majorana-Weyl spinor
$\Psi$ satisfies both $\Psi^* = \Psi$ and $\Gamma_{11} \Psi = \Psi$ where we
choose the eigenvalue for $\Gamma_{11}$ as 1. Such a spinor can be expressed
in terms of a $D = 8$ real spinor, for example the $\lambda$, as 
\begineq
\Psi = \left(\begin{array}{c}
             \lambda\\
            \Sigma^1 \lambda\end{array}\right).
\endeq

We are now ready to derive eq. (7) from eq. (5). If we write 
\begineq
\epsilon_L =  \left(\begin{array}{c}
             \lambda_L\\
            \Sigma^1 \lambda_L\end{array}\right), \qquad 
\epsilon_R = \left(\begin{array}{c}
             \lambda_R\\
            \Sigma^1 \lambda_R\end{array}\right),
\endeq
and notice that
\eqabegin
\Gamma^2 \Gamma^3 \cdots \Gamma^9 &=& \Sigma^2 \otimes \sigma_3,\nonumber\\
                                  &=& \left(\begin{array}{cc}
                 \Sigma^2 & 0\\
                 0 &- \Sigma^2\end{array}\right),
\eqaend
then eq. (5) is equivalent to the following equations
\begineq
\lambda_L = \Sigma^2 \lambda_L, \qquad \lambda_R = - \Sigma^2 \lambda_R.
\endeq
    
Note both $\lambda_L$ and $\lambda_R$ are arbitrary 16-component real spinors.
If we express them in terms of the two $D = 8$ pseudo Majorana spinors
$\epsilon^{(1)}$ and $\epsilon^{(2)}$, then the above SUSY condition should 
become the SUSY condition for the $(1,0,0)$-string configuration. Using 
eq. (28), we can identify
\eqabegin
\lambda_L &=& \frac{1}{2} (1 - \gamma^0) (\epsilon^{(1)} + \epsilon^{(2)}),
\nonumber \\
\lambda_R &=& - \frac{1}{2} (1 - \gamma^0) (\epsilon^{(1)} - \epsilon^{(2)}).
\eqaend
Substituting the above to eq. (35), we end up with eq. (7) where we have 
used $\gamma^1 = \Sigma^2$.  
 
\vfil\eject
\begin{Large}
\noindent{\bf Acknowledgements}
\end{Large}

\medskip

We are grateful to Imsoek Yang for discussion. JXL acknowledges the support
of NSF Grant PHY-9722090.

\bigskip

\begin{Large}
\noindent{\bf References}
\end{Large}

\medskip

\begin{enumerate}

\item J. H. Schwarz, {\it Lectures on Superstring and M-theory
Dualities}, hep-th/9607201.

\item O. Aharony, J. Sonnenschein and S. Yankielowicz, \np 474 (1996) 309
[hep-th/9603009].

\item M. Gaberdiel and B. Zwiebach, \np 518 (1998) 151 [hep-th/9709013].

\item K. Dasgupta and S. Mukhi, \pl 423 (1998) 261 [hep-th/9711094].

\item J. Polchinski, Phys. Rev. Lett. 75 (1995) 47; {\it TASI Lectures on
D-Branes}, hep-th/9611050.

\item A. Strominger, \pl 383 (1996) 44 [hep-th/9512059].

\item C. Callan and J. Maldacena, \np 513 (1998) 198 [hep-th/9708147].

\item O. Bergman, {\it Three-Pronged Strings and 1/4 BPS States in
$N = 4$ Super Yang-Mills Theory}, hep-th/9712211.

\item S. -J. Rey and J. -T. Yee, {\it BPS Dynamics of Triple $(p,q)$
Junctions}, hep-th/9711202.

\item C. G. Callan and L. Thorlacius, {\it Worldsheet Dynamics of String
Junctions}, hep-th/9803097.

\item M. Krough and S. Lee, {\it String Network from M-Theory}, hep-th/
9712070.

\item Y. Matsuo and K. Okuyama, {\it BPS Condition of String Junction 
from M-theory}, hep-th/9712070.

\item J. X. Lu and S. Roy, {\it U-duality p-branes in diverse dimensions},
hep-th/9805180.

\item M. Cederwall, A. von Gussich, B.E.W. Nilsson and A. Westenberg, 
\np 490 (1997) 163 [hep-th/9610148].

\item M. B. Green and M. Gutperle, \pl 377 (1996) 28 [hep-th/9602077].

\item D. Bak, J. Lee and H. Min, {\it Dynamics of BPS states in the 
Dirac-Born-Infeld theory}, hep-th/9806149.

\item E. Witten, \np 460 (1996) 335 [hep-th/9510135].

\item A. Sen,  JHEP 03 (1998) 005 [hep-th/9711130].

\item J. Schwarz, \np 226 (1983) 269.

\item T. Ortin,  Phys. Rev. D 51 (1995) 790 [hep-th/9404035].

\item S. Bhattacharyya, A. Kumar and S. Mukhopadhyay, {\it String network and
U-duality}, hep-th/9801141.

\item M. Green, J. Schwarz and E. Witten, {\it Superstring Theory}, Vol. 1, 
Cambridge Unversity Press (1987).

\end{enumerate}
\end{document}